\input{aipcheck}

\documentclass[
    ,final            
  ]
  {aipproc}

\layoutstyle{6x9}
\newcommand{\be}{\begin{equation}}
\newcommand{\ee}{\end{equation}}
\newcommand{\bea}{\begin{eqnarray}}
\newcommand{\eea}{\end{eqnarray}}

\begin{document}

\title{High mass diffraction at the LHC}

\classification{}
\keywords      {}

\author{C. Royon}{
  address={Service de physique des particules, CEA/Saclay,
  91191 Gif-sur-Yvette cedex, France \\ Fermilab, Batavia, USA}}



\begin{abstract}
We use a Monte Carlo implementation of recently developped 
models of exclusive diffractive $W$, top, Higgs and stop productions
to assess the sensitivity of the LHC experiments.\end{abstract}

\maketitle


\section{Theoretical framework}
The proposed  model for $pp \rightarrow p+H+p$, the Bialas-Landshoff (BL)
\cite{bialas} model, 
is based on a summation of two-gluon exchange Feynman graphs coupled to Higgs 
production by the top 
quark loop. 
The non-perturbative character of diffraction at the proton vertices relies 
on the introduction of 
``non-perturbative'' 
gluon propagators which are modeled on the description of soft total 
cross-sections within the additive 
constituent quark model.

More details about the theoretical model and its phenomenological
applications can be found in Refs. \cite{ourpap} and \cite{us}. In the following,
we use the BL model for exclusive Higgs production recently implemented in
a Monte-Carlo generator \cite{ourpap}. 

\section{Experimental context}

The analysis is based on a fast simulation of the CMS detector at the LHC
(similar results would be obtained using the ATLAS simulation).
The calorimetric coverage of the CMS experiment ranges up to a pseudorapidity 
of $|\eta|\sim 5$. 
The region devoted
to precision measurements lies within $|\eta|\leq 3$, with a typical 
resolution on jet energy measurement of $\sim\!50\%
/\sqrt{E}$,
where $E$ is in GeV, and a granularity in pseudorapidity and azimuth of 
$\Delta\eta\times\Delta\Phi \sim 0.1\times 0.1$. 

In addition to the central CMS detector, the existence of roman pot detectors
allowing to tag diffractively produced protons,
located on both $p$ sides, is assumed \cite{helsinki}. The $\xi$ acceptance and 
resolution have been derived for each device using a complete simulation
of the LHC beam parameters. The combined $\xi$ acceptance is $\sim 100\%
$ 
for $\xi$ ranging from $0.002$ to $0.1$, where
$\xi$ is 
the proton fractional momentum loss. The acceptance limit of the device 
closest to the interaction point
is $\xi > \xi_{min}=$0.02. 

In exclusive double Pomeron exchange, the mass of the central 
heavy object is given by $M^2 = \xi_1\xi_2 s$, where $\xi_1$ and $\xi_2$ are
the proton fractional momentum losses measured in the roman pot detectors.

\section{Results on diffractive Higgs production}
Results are given in Fig. 1 for a Higgs mass of 120 GeV, 
in terms of the signal to background 
ratio S/B, as a function of the Higgs boson mass resolution.

In order to obtain an S/B of 3 (resp. 1, 0.5), a mass resolution of about
0.3 GeV (resp. 1.2, 2.3 GeV) is needed. The forward detector design of 
\cite{helsinki} 
claims a resolution of about 2.-2.5 GeV, which leads to a S/B of about 
0.4-0.6. Improvements in this design
would increase the S/B ratio as indicated on the figure.
As usual, this number is enhanced by a large factor if one considers 
supersymmetric Higgs boson 
production with favorable Higgs or squark field mixing parameters.

The cross sections obtained after applying the survival probability of 0.03 at
the LHC as well as the S/B ratios are given in Table \ref{sb} if one assumes a
resolution on the missing mass of about 1 GeV (which is the most optimistic
scenario). The acceptances of the roman pot detectors as well as the simulation
of the CMS detectors have been taken into account in these results. 

Let us also notice that the missing mass method will allow to perform a $W$ 
mass measurement using exclusive (or quasi-exclusive) $WW$ 
events in double Pomeron exchanges, and QED processes. The advantage of the
QED processes is that their cross section is perfectly known and that this
measurement only depends on the mass resolution and the roman pot acceptance.
In the same way, it is possible to measure the mass of the top quark in
$t \bar{t}$ events in double Pomeron exchanges.

\begin{table}
\begin{tabular}{|c||c|c|c|c|c|} \hline
$M_{Higgs}$& cross & signal & backg. & S/B & $\sigma$  \\
 & section &  & & & \\
\hline\hline
120 & 3.9 & 27.1 & 28.5 & 0.95 & 5.1  \\
130 & 3.1 & 20.6 & 18.8 & 1.10 & 4.8  \\
140 & 2.0 & 12.6 & 11.7 & 1.08 & 3.7  \\ 
\hline
\end{tabular}
\caption{Exclusive Higgs production cross section for different Higgs masses,
number of signal and background events for 100 fb$^{-1}$, ratio, and number of
standard deviations ($\sigma$).}
\label{sb}
\end{table}

The diffractive SUSY Higgs boson production cross section is noticeably enhanced 
at high values of $\tan \beta$ and since we look for Higgs decaying into $b
\bar{b}$, it is possible to benefit directly from the enhancement of the cross
section contrary to the non diffractive case. A signal-over-background up to a
factor 50 can be reached for 100 fb$^{-1}$ for $\tan \beta \sim 50$
\cite{lavignac}. 


\begin{figure}[htb]
\includegraphics[width=12.5cm,clip=true]{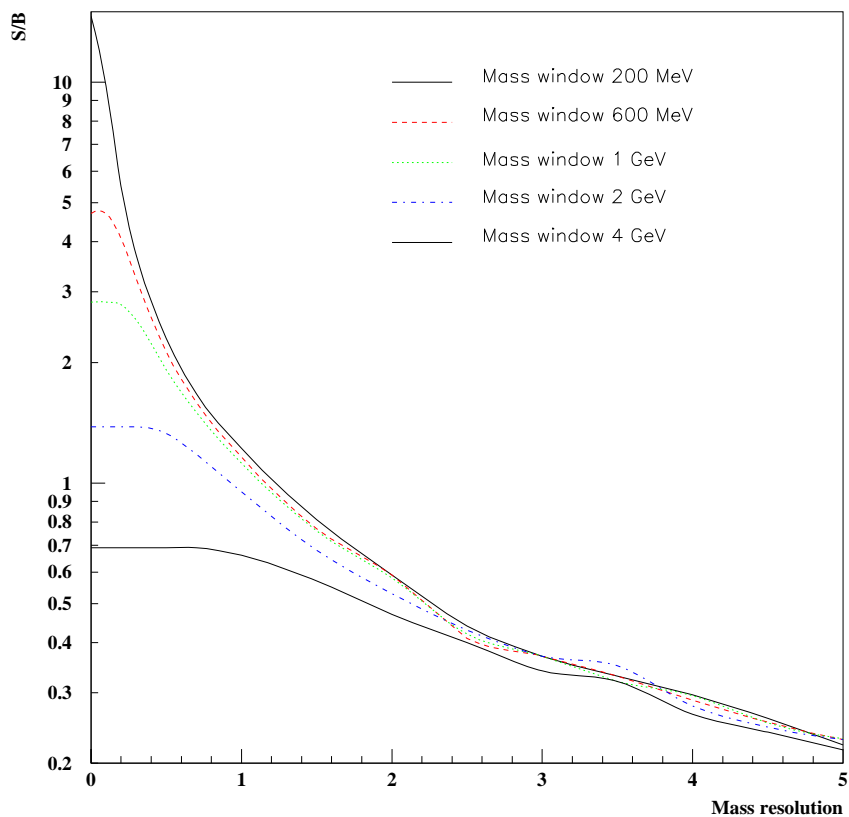}
\caption{Standard Model Higgs boson signal to background ratio as a function 
  of the resolution on
        the missing mass, in GeV. This figure assumes a Higgs
        boson mass of 120 GeV.}
\end{figure}

\section{Threshold scan method: $W$, top and stop mass measurements}
We propose a new method to measure heavy particle properties via double 
photon and double pomeron exchange (DPE), at the LHC \cite{ushiggs}. In this category of events, the heavy objects 
are produced in pairs, whereas the beam particles
often leave the interaction region intact, and can be measured using very forward detectors.

Pair production of $WW$ bosons and top quarks in QED and  double pomeron exchange are described in detail in this section. 
$WW$ pairs are produced in photon-mediated processes, which are exactly calculable in QED. There is 
basically no uncertainty concerning the possibility of measuring these processes
at the LHC. On the contrary, $t \bar{t}$ events, produced in 
exclusive double pomeron exchange, suffer from theoretical uncertainties since 
exclusive diffractive production is still to be observed at the Tevatron, 
and other models lead to different cross sections, and thus to a different
potential for the top quark mass measurement. However, since the exclusive 
kinematics are simple, the model dependence will be essentially reflected by 
a factor in the effective luminosity for such events.

\subsection{Explanation of the methods}
We study two different methods to reconstruct the mass of heavy objects
double diffractively produced at the LHC. The method is
based on a fit to the turn-on point of the missing mass distribution at 
threshold. 

One proposed method (the ``histogram'' method) corresponds to the comparison of 
the mass distribution in data with some reference distributions following
 a Monte Carlo simulation of the detector with different input masses
corresponding to the data luminosity. As an example, we can produce 
a data sample for 100 fb$^{-1}$ with a top mass of 174 GeV, and a few 
MC samples corresponding to top masses between 150 and 200 GeV by steps of. 
For each Monte Carlo sample, a $\chi^2$ value corresponding to the 
population difference in each bin between data and MC is computed. The mass point 
where
the $\chi^2$ is minimum corresponds to the mass of the produced object in data.
This method has the advantage of being easy but requires a good
simulation of the detector.

The other proposed method (the ``turn-on fit'' method) is less sensitive to the MC 
simulation of the
detectors. As mentioned earlier, the threshold scan is directly sensitive to
the mass of the diffractively produced object (in the $WW$W case for instance, it
is sensitive to twice the $WW$ mass). The idea is thus to fit the turn-on
point of the missing mass distribution which leads directly to the mass 
of the produced object, the $WW$ boson. Due to its robustness,
this method is considered as the ``default" one in the following.

\subsection{Results}

To illustrate the principle of these methods and their achievements,
we  apply them to the 
$WW$ boson and the top quark mass measurements in the
following, and obtain the reaches at the LHC. They can be applied to other 
threshold scans as well.
The precision of the $WW$ mass measurement (0.3 GeV for 300 fb$^{-1}$) is not competitive with other 
methods, but provides a very precise calibration 
of the roman pot detectors. The precision of
the top mass measurement is however competitive, with an expected precision 
better than 1 GeV at high luminosity. The resolution on the top mass is given
in Fig. 2 as a function of luminosity for different resolutions of the roman
pot detectors.

The other application is to use the so-called ``threshold-scan method"
to measure the stop mass in {\it exclusive} events. The idea is straightforward: 
one
measures the turn-on point in the missing mass distribution at about twice
the stop mass. After taking into account the stop width, we obtain a resolution
on the stop mass of 0.4, 0.7 and 4.3 GeV for a stop mass of 174.3, 210 and 393
GeV for a luminosity (divided by the signal efficiency) of 100 fb$^{-1}$. We
notice that one can expect to reach typical mass resolutions which can be obtained at a linear
collider. The process is thus similar to  those at linear colliders (all final 
states
are detected) without the initial state radiation problem. 

The caveat is  of course that production via diffractive 
{\it exclusive} processes is model dependent, and definitely needs
the Tevatron data to test the models. It will allow to determine more precisely 
the production cross section by testing and measuring at the Tevatron the jet 
and photon production for high masses and high dijet or diphoton mass fraction.

\begin{figure}[htb]
\includegraphics[width=12.5cm,clip=true]{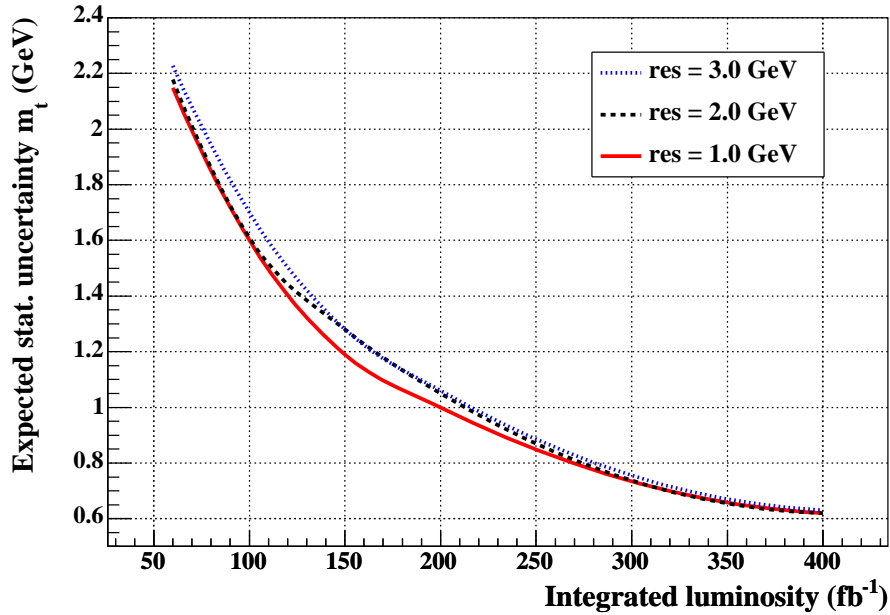}
\caption{Expected statistical precision of the top mass
    as a function of the integrated luminosity for various resolutions
    of the roman pot detectors (full line: resolution of 1 GeV, dashed line: 2
    GeV, dotted line: 3 GeV).}
\end{figure}

\section*{Acknowledgments}
There results come from a fruitful collaboration with M. Boonekamp, J. Cammin,
S. Lavignac and R. Peschanski.



 



\end{document}